\documentclass{eage2020}
\usepackage{mathtools}
\usepackage{amsmath}
\usepackage{amsfonts}
\usepackage{amssymb}
\usepackage{color}
\begin{document}
  
\title{Iterative frequency-domain seismic wave solvers based on multi-level domain-decomposition preconditioners}

\author{
        V. Dolean\footnotemark[1]$~$                                  
        P.~Jolivet\footnotemark[2], $~$
        P.-H.~Tournier\footnotemark[3] $~$
        S.~Operto\footnotemark[4]
         \\
\footnotemark[1] $~$ Univ. of Strath/UCA-LJAD;
\footnotemark[2] $~$ IRIT-CNRS;
\footnotemark[3] $~$ University Paris 6;
\footnotemark[4] $~$ UCA-Geoazur;
        }
\maketitle

\section{Summary}
Frequency-domain full-waveform inversion (FWI) is suitable for long-offset stationary-recording acquisition, since reliable subsurface models can be reconstructed with a few frequencies and attenuation is easily implemented without computational overhead. In the frequency domain, wave modelling is a Helmholtz-type boundary-value problem which requires to solve a large and sparse system of linear equations per frequency with multiple right-hand sides (sources). This system can be solved with direct or iterative methods. While the former are suitable for FWI application on 3D dense OBC acquisitions covering spatial domains of moderate size, the later should be the approach of choice for sparse node acquisitions covering large domains (more than 50 millions of unknowns). Fast convergence of iterative solvers for Helmholtz problems remains however challenging due to the non definiteness of the Helmholtz operator, hence requiring efficient preconditioners. In this study, we use the Krylov subspace GMRES iterative solver combined with a multi-level domain-decomposition preconditioner. Discretization relies on continuous finite elements on unstructured tetrahedral meshes to comply with complex geometries and adapt the size of the elements to the local wavelength ($h$-adaptivity). We assess the convergence and the scalability of our method with the acoustic 3D SEG/EAGE Overthrust model up to a frequency of 20~Hz and discuss its efficiency for multi right-hand side processing.

\newpage
\clearpage

\section{Introduction}
\vspace{-0.3cm}
The ocean bottom node (OBN) acquisition is emerging for deep-offshore seismic imaging by full waveform inversion (FWI) \citep{Beaudoin_2007_FDO}.
These stationary-recording acquisitions have the versatility to design ultra-long offset surveys, which provide a wide angular illumination of the subsurface amenable to broadband velocity models.
This wide-angle illumination allows for efficient frequency-domain (FD) FWI by decimating the multi-fold wavenumber coverage through a coarse frequency sampling  \citep{Pratt_1999_SWIb}. This frequency subsampling makes FD modelling competitive with time-marching methods and leads to compact datasets \citep{Plessix_2017_CAT}. 
Moreover, attenuation is easily implemented in FWI without computational overheads, even improving the conditioning of Helmholtz operators. \\
%
In this context, we present a new solver for 3D FD wave simulation as a forward engine for FWI. 
FD seismic modelling is a boundary-value problem, which requires to solve a sparse linear system whose unknown is the wavefield, the right-hand side (RHS) the seismic source and the coefficients embed the subsurface properties. Two main linear algebra methods exist to solve such a system. The first relies on {\it{sparse direct solver}} \citep{Duff_2017_DMS} with the advantages of accurate solutions in a finite number of operations and efficient processing of multiple RHSs for problems of moderate size (< $50.10^6$ unknowns) \citep{Amestoy_2016_FFF,Mary_2017_PHD}. The pitfalls are the memory overhead generated by the storage of the LU factors and the limited scalability of the LU decomposition, which makes application on large scale problems challenging (> $50.10^6$ unknowns). The second approach relies on {\it{iterative solvers}} \citep{Saad_2003_IMS}, whose natural scalability and moderate memory demand make them suitable for large-scale problems. However, two issues are the convergence speed of iterative solvers for ill-conditioned Helmholtz problems, which critically depend on preconditioning with the ultimate goal to make the iteration count independent to frequencies, and the efficient processing of multiple RHSs. \\
Here, we focus on the second category because we target large computational domains (several hundred of millions of unknowns) with a limited number of reciprocal sources (from few hundreds to few thousands). Our method relies on a finite-element discretization on a tetrahedral mesh, the Krylov subspace GMRES solver \citep{Saad_2003_IMS} and a Schwarz multi-level domain decomposition preconditioner \citep{Graham_2017_RRD}. Compared to the celebrated preconditioner based upon shifted Laplacian and multigrid method \citep{Erlangga_2008_ETNA}, it is less sensitive to the shift (added attenuation) and can be used without it.
In the following, we briefly review the method, before assessing the strong and weak scalability of the solver on the 3D SEG/EAGE Overthrust model. 

\section{Iterative solver for the Helmholtz problem with a domain decomposition preconditioner}
\vspace{-0.3cm}
We seek to develop a robust preconditioned iterative solver for the Helmholtz equation
\begin{equation}
\left(\Delta + k^2(\bold{x}) \right) u(\bold{x},\omega) = b(\bold{x},\omega), ~ \text{in a subsurface domain } \Omega,
\label{eqh}
\end{equation}
where $u$ is the monochromatic pressure wavefield, $b$ the monochromatic source, $k(\bold{x},\omega)=\omega/c(\bold{x})$, with $\omega$ denoting frequency, $c(\bold{x})$ the wavespeed (which is complex valued in viscous media) and $\bold{x}=(x,y,z) \in \Omega$. 
Equation~\eqref{eqh} is implemented with absorbing boundary conditions along the vertical and bottom faces of $\Omega$ and a homogeneous Dirichlet condition on the pressure along the top face. \\
We discretize Equation~\eqref{eqh} with Lagrange finite elements of degree 2 (P2) on a tetrahedral mesh $\Gamma$ of the domain  $\Omega$, leading to the following linear system 
\begin{equation}
\bold{A} \bold{u} = \bold{b}.
\label{eq1}
\end{equation}
A well-known iterative solver for this type of indefinite linear systems is the Krylov subspace Generalized Minimal RESidual Method (GMRES) \citep{Saad_2003_IMS}. However, the Helmholtz operator requires efficient preconditioning \citep[][ section 2.2.1]{Dolean_2015_IDD}. \\
In this study, we solve system \eqref{eq1} with a two-level domain decomposition preconditioner $\bold{M}^{-1}$
\begin{equation}
\label{2lvl}
\bold{M}^{-1}  = \bold{M}^{-1}_1 (I - \bold{A} \bold{Q}) + \bold{Q}, \quad \text{with } \bold{Q} = \bold{Z} \bold{E}^{-1} \bold{Z}^T, \quad \bold{E} = \bold{Z}^T \bold{A} \bold{Z},  \\
\end{equation}
where $\bold{M}^{-1}_1$ is the one-level Optimized Restricted Additive Schwarz (ORAS) preconditioner and $\bold{Z}^T$ is the interpolation matrix from the finite element space defined on $\Gamma$ onto a finite element space defined on a coarse mesh $\Gamma_H$ \citep{Bonazzoli_2018_TDP}.
The construction of the domain decomposition preconditioner is described in detail in \citet{Bonazzoli_2018_TDP}. Let $\left\{\Gamma_i\right\}_{1 \le i \le N_d}$ be an overlapping decomposition of the mesh $\Gamma$ into $N_d$ subdomains. Let $\left\{\bold{A}_i\right\}_{1 \le i \le N_d}$ denote local Helmholtz operators with absorbing (or transmission) boundary conditions at the subdomain interfaces. The one-level ORAS preconditioner is
\begin{equation}
\bold{M}^{-1}_1 = \sum_{i=1}^{N_d} \bold{R}_i^T \bold{D}_i \bold{A}_i^{-1} \bold{R}_i,
\label{oras}
\end{equation}
where $\left\{\bold{R}_i\right\}_{1 \le i \le N_d}$ are the Boolean restriction matrices from the global to the local finite element spaces and $\left\{\bold{D}_i\right\}_{1 \le i \le N_d}$ are local diagonal matrices representing the partition of unity.\\
The key ingredient of the ORAS method is that the local matrices $\left\{\bold{A}_i\right\}_{1 \le i \le N_d}$ incorporate more efficient boundary conditions (i.e. absorbing boundary conditions) than in the standard RAS preconditioner based on local Dirichlet boundary value problems.\\
The scalability is achieved by the iterative solution of the coarse problem $\bold{E}$ in~\eqref{2lvl} again using GMRES with a one-level ORAS preconditioner. We use the same spatial subdomain partitioning for the coarse and fine meshes. Each computing core is assigned to one spatial subdomain and holds the corresponding coarse and fine local matrices. Each application of the global preconditioner $\bold{M}^{-1}$ relies on local concurrent subdomain solves on the coarse and fine levels, which are performed by a direct solver. This hybrid direct/iterative solver requires careful strong scalability analysis to achieve the best compromise between parallel efficiency and memory storage.\\
When processing multiple RHSs, the application of the preconditioner exploits the multi-RHS capabilities of the direct solver for the local forward elimination and backward substitution in each subdomain.

\section{Numerical experiments}
\vspace{-0.3cm}

\begin{figure}[ht!]
\begin{center}
\includegraphics[width=8cm,clip=true,trim=0cm 5cm 0cm 0cm]{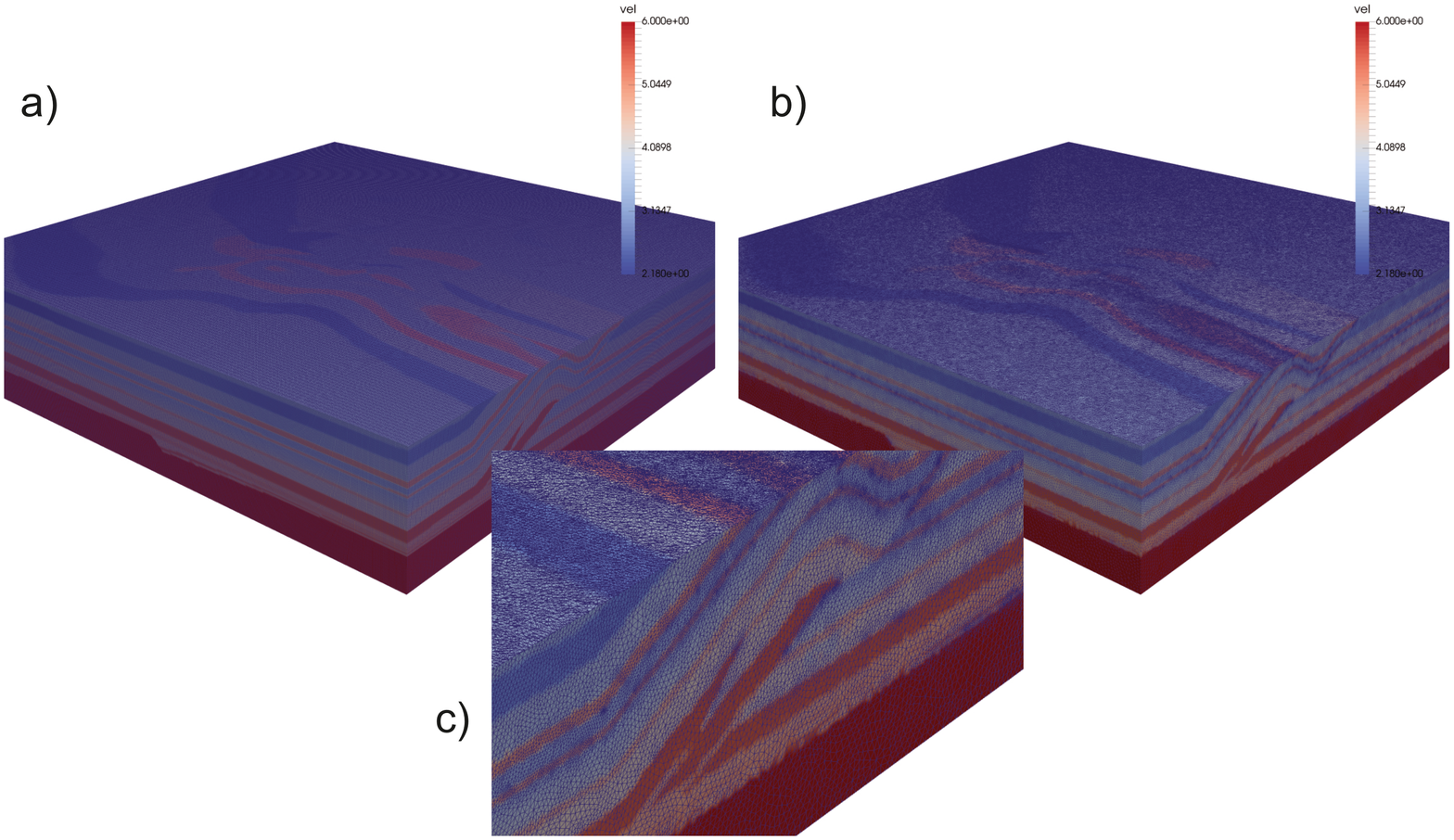}
\caption{3D SEG/EAGE Overthrust model. (a) Cartesian. (b,c) Tetrahedral mesh.}
\label{over1}
\end{center}
\end{figure}

\begin{figure}[ht!]
\begin{center}
\includegraphics[width=11cm,clip=true,trim=0cm 5cm 0cm 0cm]{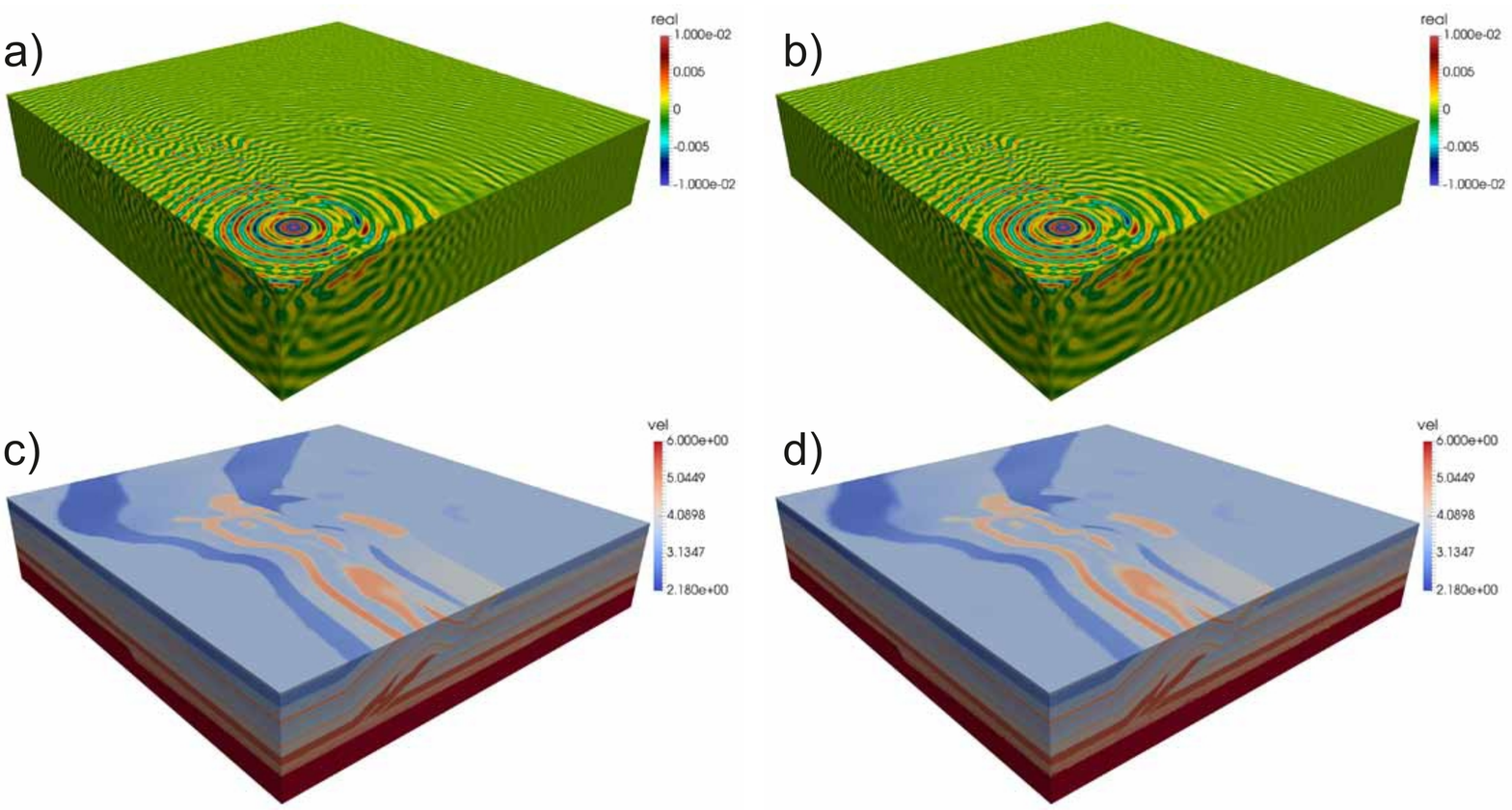}
\caption{10~Hz monochromatic wavefields. (a) Cartesian. (b) Tetrahedral meshes.}
\label{over2}
\end{center}
\end{figure}

\begin{figure}[ht!]
\begin{center}
\includegraphics[width=10cm,clip=true,trim=0cm 8cm 0cm 0cm]{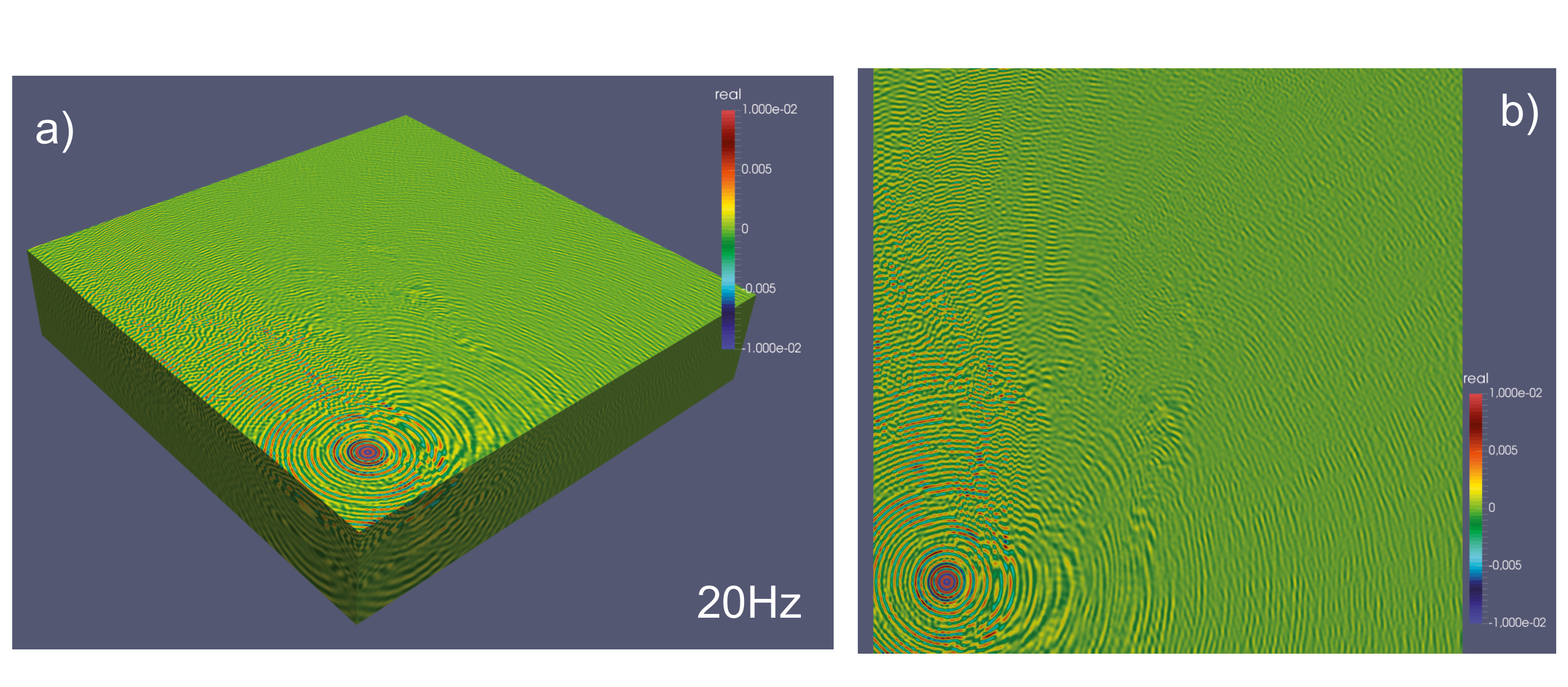}
\caption{20~Hz monochromatic wavefield in the tetrahedral mesh.}
\label{over3}
\end{center}
\end{figure}

\begin{figure}[ht!]
\begin{center}
\includegraphics[height=4.5cm,clip=true,trim=0cm 0cm 0cm 0.1cm]{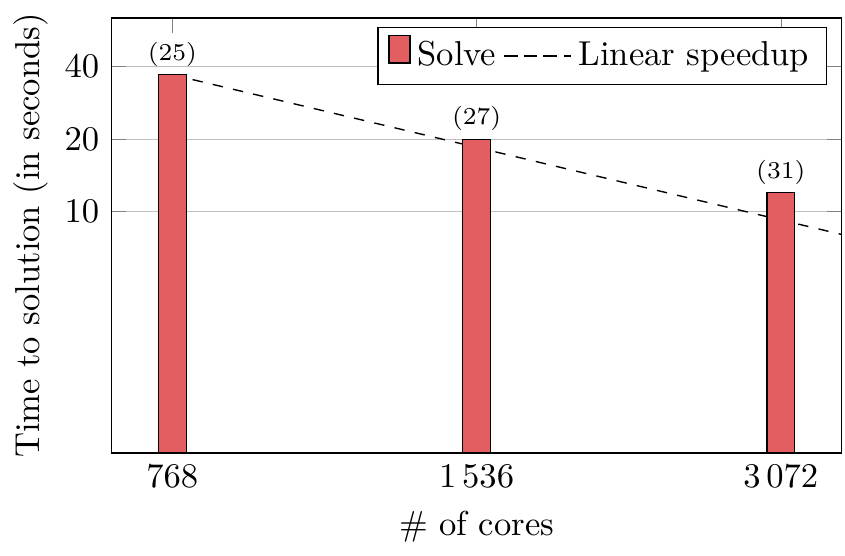}
\includegraphics[height=4.5cm,clip=true,trim=0cm 0cm 0cm 0cm]{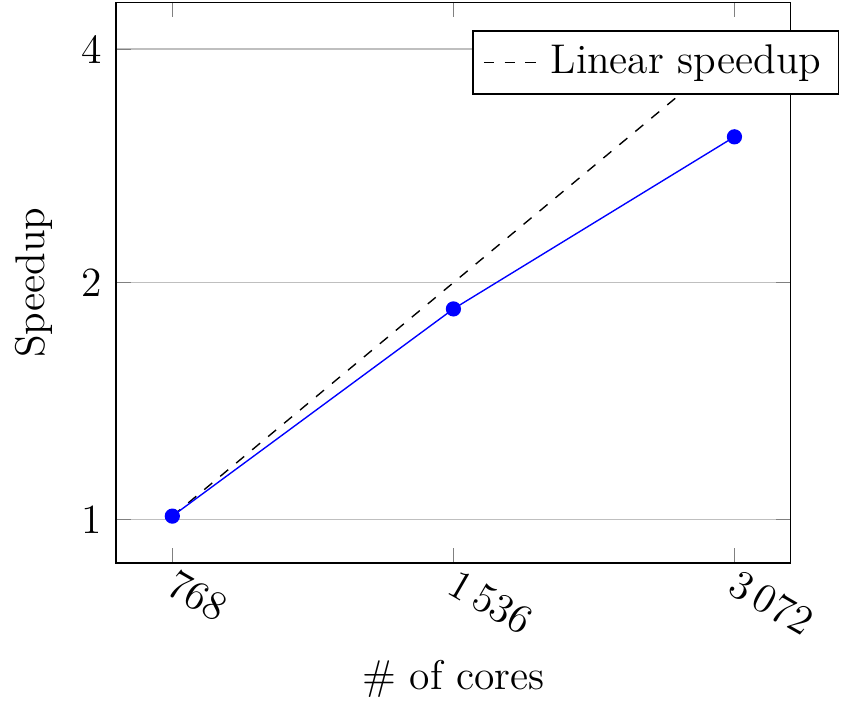}
\vspace{-0.4cm}
\caption{Strong scaling (10~Hz) on adaptive tetrahedral mesh. Number of iterations in brackets.}
\label{over4}
\end{center}
\end{figure}

The solver is implemented using the high-performance domain decomposition library HPDDM (High-Performance unified framework for Domain Decomposition Methods, http://github.com/hpddm/hpddm)  \citep{Jolivet_SC13}.
We assess the solver on the Occigen supercomputer of CINES (https://www.cines.fr) with the 3D $20 \times 20 \times 4.65$ km SEG/EAGE Overthrust model (Fig.~\ref{over1}). We perform wave simulation  with P2 finite elements on Cartesian and adaptive tetrahedral meshes for the 5~Hz, 10~Hz and 20~Hz frequencies (\textcolor{black}{Tab.}~\ref{tab_over}).
 The average length of the element edges is set to 5 nodes per minimum wavelength on the Cartesian grid, and  5 nodes per local wavelengths in the tetrahedral mesh (2.5 for the coarser mesh used in the two-level method).
We use a homogeneous Dirichlet boundary condition at the surface and first-order absorbing boundary conditions along the other five faces of the model. The source is located at (2.5,2.5,0.58) km.  For weak scalability analysis, we keep $\#$dofs per subdomain roughly constant from one frequency to the next (\textcolor{black}{Tab.}~\ref{tab_over}). The $h$-adaptivity in the tetrahedral mesh decreases $\#$dofs relative to the Cartesian mesh by a factor of 2.07. The stopping tolerance for GMRES is that the relative residual $\|\bold{Ax}-\bold{b}\|/\|\bold{b}\|$ is reduced by $10^{-6}$. 
he consistency between the 10~Hz wavefields computed in the Cartesian and tetrahedral meshes is shown in Fig.~\ref{over2}. Timings for the tetrahedral mesh are more than three times smaller than those obtained on the Cartesian mesh (\textcolor{black}{Tab.}~\ref{tab_over}). The simulation at 20~Hz on the adaptive mesh (Fig.~\ref{over3}) involves 678 millions of dofs and requires 12,288 cores (512 computer nodes equipped with 2 CPUs and 12 cores per CPU). For the 1-level preconditioner, the number of iterations roughly linearly increases with frequency, which is consistent with the results of \citet{Plessix_2017_CAT} based on iterative solver preconditioned with a multi-grid technique and shifted Laplacian. However, the 2-level preconditioner outperforms significantly the one-level method in every test case. On the tetrahedral mesh, the elapsed time achieved by the 2-level preconditioner is  20s and 90s for 10~Hz and 20~Hz respectively, while it is 93s and 298s for the 1-level counterpart  (\textcolor{black}{Tab.}~\ref{tab_over}). The strong scalability of the solver is shown in Fig.~\ref{over4} for the 10~Hz frequency.
When dealing with multiple RHSs, the pseudo-block GMRES yields a speedup of around 2 for blocks of 10 RHSs. 
Finally, with physical attenuation, the iteration count decreases as expected. For example, with constant $Q = 200$, the iteration count at 10~Hz with the tetrahedral mesh decreases from 27 to 19 with 1,536 subdomains and the computing time decreases from 20s to 14s.
%
%
%
%
\vspace{-0.5cm}
\vspace{-0.3cm}
\section{Conclusions}
\vspace{-0.3cm}
We have proposed a massively-parallel iterative solver as a forward engine for 3D frequency-domain FWI from ultra-long offset stationary-recording survey.
Multi-RHS processing can be further improved with block and recycling strategies, in particular the Krylov subspace recycling method GCRO-DR \citep{Parks_2006_RKS} and its block variant, which are already implemented in the HPDDM library and have been applied successfully for medical imaging based on a multi-antenna microwave device \citep{Jolivet_2016_BIM}. Extension to visco-elastic media is also scheduled.

%

%
%

%
\begin{table}[ht!]
\begin{center}
\begin{tabular}{|c|c|c|c|c|c|c|c|c|c|}
\hline
\multicolumn{4}{|c|}{\bf{Cartesian grid}} & \multicolumn{2}{c|}{\bf{1-level}} & \multicolumn{3}{c|}{\bf{2-level}}\\ \hline
Freq (Hz) & $\#$core & $\#$elts  (M)& $\#$dofs (M)   & $\#$it & 1 RHS & $\#$it & 1 RHS & 10 RHSs  \\ \hline
{\bf{5}} & 192 & 16 & 22 & 117 & 109s & 18 & 28s & OOM \\ \hline
{\bf{10}} & 1,536 & 131 & 176 & 249 & 242s & 45 & 75s & OOM  \\ \hline
{\bf{20}} & 12,288 & 1048 & 1408 & 506 & 998s & 117 & 396s & OOM  \\ \hline
\multicolumn{4}{|c|}{\bf{Tetrahedral mesh}} & \multicolumn{2}{c|}{\bf{1-level}} & \multicolumn{3}{c|}{\bf{2-level}}\\ \hline
Freq (Hz) & $\#$core & $\#$elts  (M)& $\#$dofs (M)   & $\#$it & 1 RHS & $\#$it & 1 RHS & 10 RHSs  \\ \hline
{\bf{5}} & 192 & 8 & 11 & 123 & 32s & / & / & /  \\ \hline
{\bf{10}} & 1,536 & 63 & 85 & 221 & 93s & 27 & 20s & 86s \\ \hline
{\bf{20}} & 12,288 & 506 & 678 & 528 & 298s & 68 & 90s & 408s \\ \hline
\end{tabular}
\end{center}
\caption{Statistics of the simulation in Cartesian and tetrahedral meshes. $Freq (Hz)$: frequency; $\#$core: number of cores; $\#$elts: number of finite elements; $\#$dofs: number of degrees of freedom; $\#$it: number of iterations. Elapsed time in seconds for 1 and 10 RHSs.}
\label{tab_over}
\end{table}

\vspace{-0.3cm}
\section{Acknowledgements}
\vspace{-0.3cm}
This study was granted access to the HPC resources of SIGAMM (http://crimson.oca.eu) and CINES/IDRIS under the allocation 0596 made by GENCI. This study was partially funded by the WIND consortium (\textit{https://www.geoazur.fr/WIND}), sponsored by Chevron, Shell and Total.
\bibliographystyle{apalike}
\newcommand{\SortNoop}[1]{}

%
%

\end{document}